\documentclass[[final,3p,times,twocolumn]{elsarticle}
\usepackage{graphicx}% Include figure files
\usepackage{dcolumn}% Align table columns on decimal point
\usepackage{bm}% bold math
\usepackage{amsmath}
\usepackage{comment}
\usepackage{epstopdf} %converting to PDF
\usepackage{xr-hyper}
\usepackage{lineno,hyperref}
\usepackage{color} %for the comments by Jo
\usepackage{soul} %Abner's comments in highlighted text
\usepackage{float}

\makeatletter
\newcommand*{\addFileDependency}[1]{% argument=file name and extension
	\typeout{(#1)}
	\@addtofilelist{#1}
	\IfFileExists{#1}{}{\typeout{No file #1.}}
}
\makeatother

\usepackage{amssymb}

\usepackage{lineno,hyperref}
\modulolinenumbers[5]

\journal{Ultramicroscopy}

\begin{document}

\begin{frontmatter}



\title{Reliable phase quantification in focused probe electron ptychography of thin materials}

\author[emat]{Christoph Hofer}
\author[emat]{Timothy J. Pennycook\corref{mycorrespondingauthor}}
\cortext[mycorrespondingauthor]{Corresponding author}
\ead{timothy.pennycook@uantwerpen.be}

\address[emat]{EMAT, University of Antwerp, Groenenborgerlaan 171, 2020 Antwerp, Belgium}
\begin{abstract}
 Electron ptychography provides highly sensitive, dose efficient phase images which can be corrected for aberrations after the data has been acquired. This is crucial when very precise quantification is required, such as with sensitivity to charge transfer due to bonding. Drift can now be essentially eliminated as a major impediment to focused probe ptychography, which benefits from the availability of easily interpretable simultaneous Z-contrast imaging. However challenges have remained when quantifying the ptychographic phases of atomic sites. The phase response of a single atom has a negative halo which can cause atoms to reduce in phase when brought closer together. When unaccounted for, as in integrating methods of quantification, this effect can completely obscure the effects of charge transfer. Here we provide a new method of quantification that overcomes this challenge, at least for 2D materials, and is robust to experimental parameters such as noise, sample tilt.
 %An advantage of focused probe ptychography is the availability of simultaneously acquired Z-contrast imaging, which remains useful for elemental identification. 
% Now that cameras are capable of collecting 4D scanning transmission electron microscopy (4D STEM) data without reducing scan speed, focused probe ptychography can be performed without significant drift. However quantifying the phases of atomic sites remains a challenge as ptychographic image contrast does not follow a simple Z dependence but is highly non-linear with a more complex dependence on the local atomic configuration. In par
 
% as well as the electronic distribution. Here, we address this issue with a new method of quantifying the atomic phases which is robust to experimental parameters such as noise, sample tilt and works very well for weak-phase objects such as 2D materials. Our method can be crucial to correctly quantifying small phase changes that occur for example with charge transfer due to bonding.
\end{abstract}

% \includegraphics{figs/grabs.pdf}

%\item Ptychographic phase contrast images are sensitive to the projected atomic distances
%\item This non-linearity makes quantification difficult which is important for e.g.\ charge transfer measurements
%\item A new method for phase quantification is introduced which is accurate and robust to experimental parameters and materials inhomogeneities.

% Keywords

\end{frontmatter}

\section{Introduction}
Electron ptychography addresses the phase retrieval problem in transmission electron microscopy~\cite{ptycho1}, using the probe position dependence of the convergent beam electron diffraction. Ptychography takes advantage of redundancy in the diffraction data, either via overlapping probe positions or overlapping diffracted beams. Before the success of aberration correction in hardware, direct focused probe ptychography was successfully applied to resolve the atomic structure at a resolution three times the conventional limit a scanning transmission electron microscope (STEM) with its ability to achieve superresolution~\cite{Nellist1995}. With the rise of aberration correction in hardware, atomic resolution became largely routine for many materials, and the limits of electron microscopy shifted more to the robustness of the sample to the electron beam in many cases. However ptychography is useful for much more than superresolution, and a primary feature of the technique is its very high dose efficiency, which can even exceed the dose-efficiency of aberration-corrected high-resolution transmission electron microscopy (HR-TEM)~\cite{PENNYCOOK2019}. This efficiency is of great interest for beam sensitive samples, but also when  the need for extreme sensitivity and precision arises, as is the case for accurate measurements of charge density, especially when one wishes to detect the subtle effect of charge transfer due to bonding.
\begin{figure*}[h]
 \centering\includegraphics[width=0.95\textwidth]{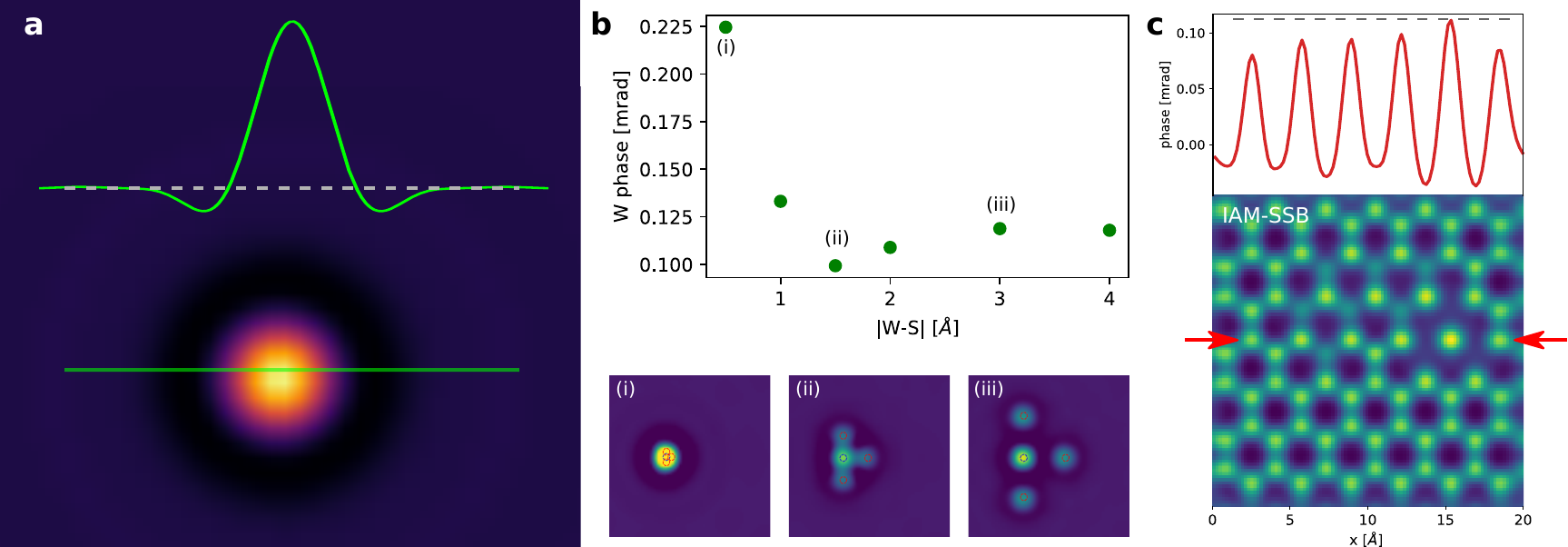}%
 \caption{\label{distance} \textbf{Dependency of SSB ptychographic images on the distance to neighboring atoms.} a) SSB image of isolated W with a line profile in green showing the negative halo. The dashed lines in the plot of the line profile indicate 0 phase.  b) Examining the SSB phase variation of neighboring single atoms as a function of their distance. The plot shows the phase at the center of a W atom vs the distance to three neighboring S atoms. The S atoms are moved outwards from the W atom as indicated in the SSB images at the bottom, using a green circle to indicate the W site and red circles to label the S atoms. Because of the negative halo, the phase at the center of W atom displays a minimum when the negative halos from the atomic sites coinside. c) This is important to take into account SSB images such as this simulation of defective WS\textsubscript{2}, which shows a significant variation of the W profile depending on positions of the neighboring atoms.}
 \end{figure*}
 
To overcome the scan speed limitations imposed by slow cameras, electron ptychography can be performed using a defocused probe setup, which allows one to utilize far fewer probe positions for a given area~\cite{Chen2020,Song2019}. This can be useful for low dose applications, as the electrons are then distributed over a larger area~\cite{Zhou2020}. Now cameras have exceeded MHz readout times allowing one to capture truly fast 4D data sets with a focused, atomically sharp probe~\cite{JANNIS2022}. This allows one to avoid both the high doses and distortions due to drift associated with slow scans with a focused probe and allows one to simultaneously capture the higher angle scattering for annular dark field (ADF) Z-contrast imaging, which would be out of focus with a defocused probe. This is an extremely useful configuration as the phase images provide the greatest sensitivity to all the atoms, and indeed electrons, while the Z-contrast ADF image remains very useful for its ease of interpretation and ability to distinguish heavier elements~\cite{Yang2016,Chuang}.

Phase images are sensitive to the total potential in the material, including the contribution from electrons in addition to the nuclear potential, enabling charge density mapping~\cite{MADSEN2021,MULLERCASPARY2017,Gao2019} and even detection of charge transfer~\cite{Martinez2019,meyer2011}. However, detecting charge transfer is particularly challenging and requires a robust and reliable method to quantify the phases of 
atomic sites. Numerous publications successfully address the quantification of atomic intensities with the ADF signal~\cite{MARTINEZ2018,Krivanek2010,E2013}. Phase contrast images are, however, significantly more complicated to analyse as the contrast transfer function is generally more complex~\cite{OLEARY2021,YANG20152}. 

 \begin{figure*}[h]
 \center\includegraphics[width=0.95\textwidth]{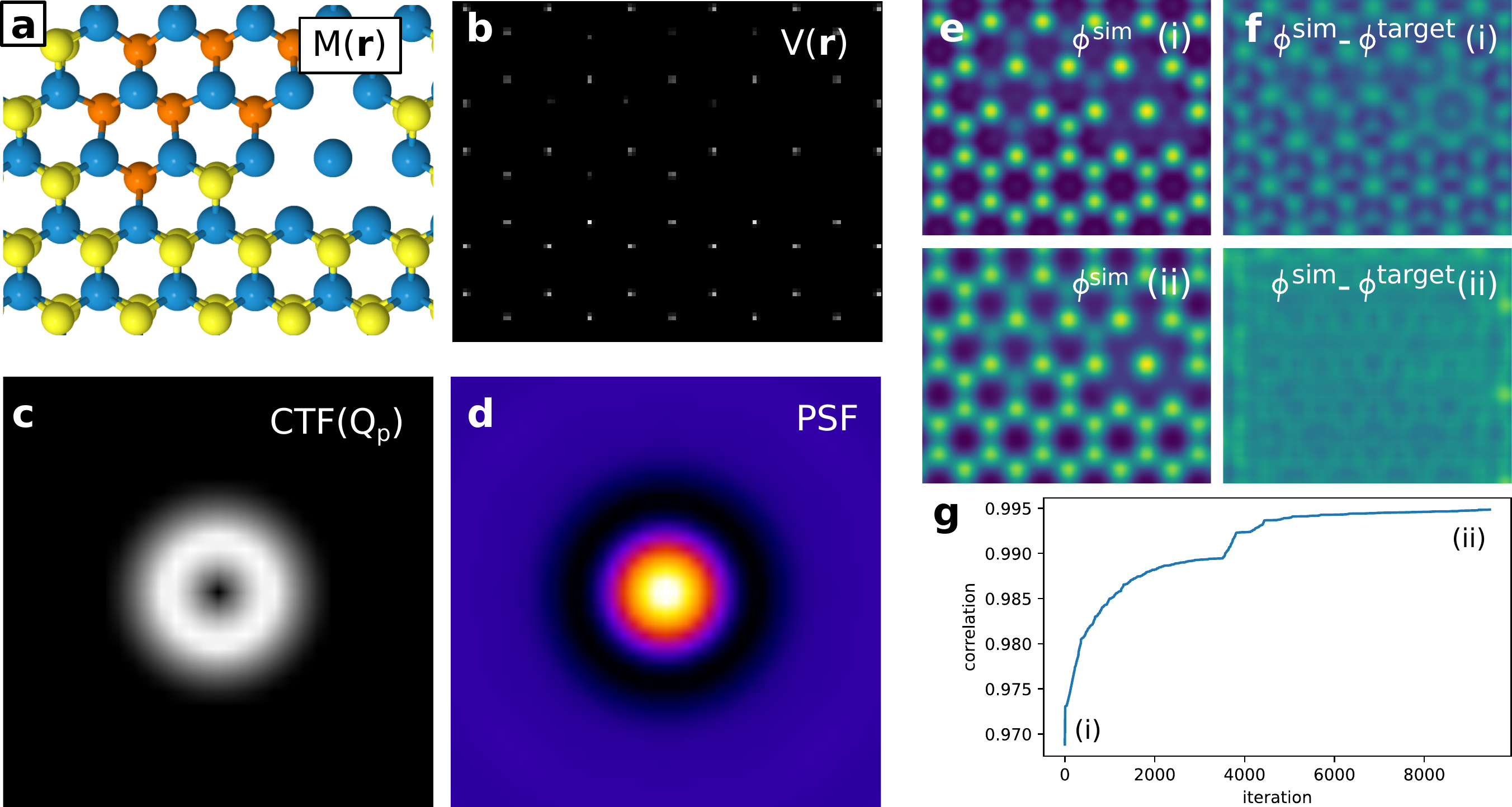}%
 \caption{\label{progress} \textbf{Illustration of the steps in the procedure for phase quantification.} a) The initial atomic model containing the correct elements and the approximate positions is converted to b) a point-potential based on the model. c) The calculated CTF of SSB ptychography is converted to the PSF by FFT as illustrated in d). e) Initial (i) and optimized (ii) convolutional simulations. As can be seen in the difference images in f), showing the convolutional simulations  minus the target image, the difference has been greatly reduced from the initial result (i) in the optimized image (ii). One can also see how the optimization increases the correlation function as a function of iteration in g).}
 \end{figure*}
Single side band (SSB) ptychography is a fast, direct phase retrieval method in STEM, extracting the phase by integrating over the double disk overlaps in probe reciprocal space~\cite{PENNYCOOK2015}. Problems quantifying SSB, or other ptychographic images, are revealed when analysing phase images of single atoms, such as those shown in Fig.~\ref{distance}. Unlike ADF images, an SSB phase image displays a negative halo around a single atom, before converging to a zero valued background. This negative contribution makes quantification difficult, as the negative tail can influence the phase of neighboring atoms. Fig.~\ref{distance}b shows the phase at the center of a single W atom as a function of the distance to three neighboring S atoms. At small distances, the atomic phases significantly overlap and add up to a large integrated phase. This occurs with any imaging technique as the distance of the atoms are smaller than the resolution limit of the microscopic setup. Increasing the distance decreases the W phase and separates the atoms. Importantly, however, the phase reaches a minimum and increases again before reaching a plateau corresponding to the phase of an isolated independent atom. The occurrence of a local minimum is a result of the negative halo of the atomic phase.

The negative halo can cause the phase of the same type of atom to vary significantly depending on the configuration of the atoms surrounding it, which can hamper quantification. To demonstrate this, we simulated an SSB image of a defective area of monolayer WS\textsubscript{2}, introducing several S vacancies to change the local environment of certain W atoms. This is shown in Fig.~\ref{distance}c. We also included a quasi isolated W atom surrounded by 3 S divacancies, which is not physically stable but is a useful demonstration of the difficulty of phase quantification. A line profile across the W atoms shows a clear variation of the observed phases. The highest phase is the quasi-isolated W site while the lowest W phase is the pristine site without vacancies. This is because the quasi-isolated W atom is not influenced by the negative contribution of neighboring atoms. The W phases within the pristine lattice have a similar distance as the minimum in Fig.~\ref{distance}b resulting in the lower phase observed. Therefore, the phase of each atom in a lattice is visually reduced compared to an isolated atom. Such effects are not accounted for using simple integration methods of quantification. Note that this is in contrast to Z-contrast ADF imaging, where no negative halos exist, only positive definite probe tails. This is important when, for instance, the sample tilts and the projected interatomic distances are changed, altering the phase of the atomic sites without meaningful change in the potential itself. 

\begin{figure}[h]
 \center\includegraphics[width=0.45\textwidth]{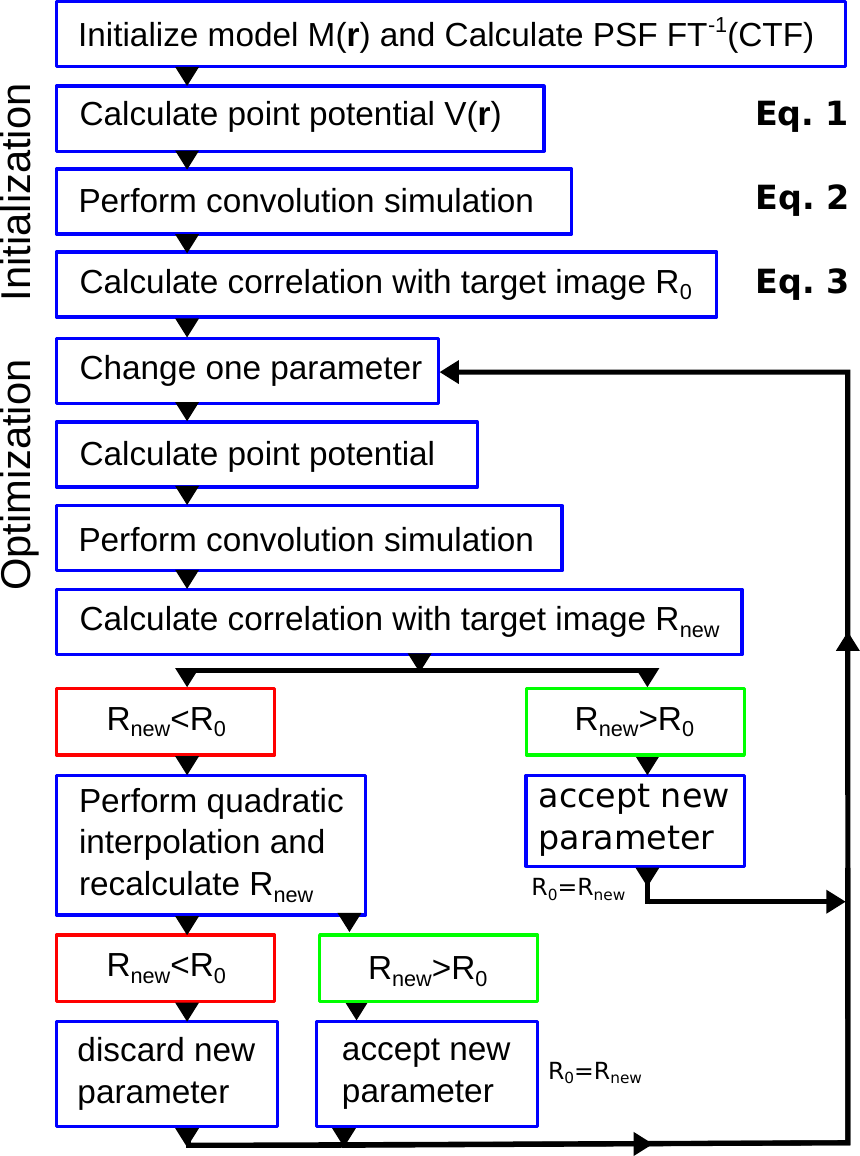}%
 \caption{\label{alg} \textbf{Flow chart of the phase extraction algorithm.}}
 \end{figure}

To account for this non-linear contrast formation, we introduce a new method which reliably and accurately extracts the phase corresponding to the actual potential for the atoms in arbitrary networks and orientations, and demonstrate it with defective WS\textsubscript{2} as shown Fig.~\ref{distance}c. The method optimizes a point-potential which is convolved with a kernel. Here we focus on the contrast transfer function (CTF) for SSB, although other kernels can be used for other imaging modes, as we demonstrate for ePIE and iCoM. The SSB kernel has the same shape as an isolated SSB reconstructed atom such as in Fig.~\ref{distance}a and can be interpreted as the point spread function (PSF). The point-potential represents the atomic positions as well as their phase. We compare this new ``SSB-CTF based'' kernel with a kernel generated by the ``ADF-CTF'' which is very similar to a Gaussian kernel. The analysis with the ADF-CTF based kernel is very similar to simple Gaussian-fitting~\cite{VANAERT2009} or Voronoi-integration~\cite{E2013} in the absence of aberrations~\cite{HOFER2021}.
The new method reliably and accurately extracts the deconvolved phases and, importantly, is robust to environmental changes such as local atomic configuration and sample tilt. Our new method in principle makes quantitative analysis of phase images with arbitrary contrast transfer functions possible.

\section{Methods}
\subsection{Computational method}
The method works by iteratively updating the atomic positions and their point potential representations such that images simulated using the convolution of the point potentials by the PSF match the input images as closely as possible. Figure \ref{progress} illustrates the process using defective WS\textsubscript{2} as an example, and the complete method is summarized in a flow chart in Fig~\ref{alg}). 
The WS\textsubscript{2} model in Fig.~\ref{progress}a results in the point-potential image shown in b. 

In general, the 2D point potential takes the form
\begin{equation}
V(\textbf{r}) = \sum_{\textbf{r}_Z}\Phi_{\textbf{r}_Z}\cdot\delta_{\textbf{r},\textbf{r}_Z}
\end{equation}
where $\textbf{r}_Z$ are the exact positions of the atoms in the model $M$ and $\delta$ is the Kronecker delta function, being one if $\textbf{r}=\textbf{r}_Z$ and zero otherwise. We chose $\Phi$ as the variable for non-zero pixel values as they represent the phase cross-section of the atoms. In practise, the potential is calculated by iterating through all atoms and adding the corresponding phase value at the corresponding pixel position.  For spatial precision, sub-pixel precision is enabled by weighting four neighboring pixels in the potential map, as the potential maps have the sample pixel dimensions as the input images, which can be courser than the positions of the atoms. In practise, the initial model is often relatively easy to match to the experimental data manually via scaling and translations.

Once a point potential map is produced, a phase image, $\phi$, is simulated simply by convolving the potential with the chosen PSF~\cite{Kirkland2010}, the inverse Fourier transform of the CTF,  
\begin{equation}
\phi^{sim}= V(\textbf{r})\ast FT^{-1}\left(CTF(Q_p)\right).
\end{equation}
The error is then calculated between the convolutionally simulated image, $\phi^\mathrm{sim}$, and the input phase image, $\phi^\mathrm{target}$, and the positions of the atoms in the model and their point potential values is then iteratively updated such that the difference between the image simulated from the point potential map and the input image is minimized. In addition the algorithm additionally optimizes the PSF width, which can be altered by either the convergence angle or the field of view via the probe position step size. The optimization process is described in detail below. Using convolutional simulations is advantageous for the speed of the computations, but by optimizing a single value of the potential for each atomic site in this way, the point is also to provide the best possible quantification of ``the phase'' of an atom or atomic column, including the effect of screening by electrons which are distributed further away from the nucleus\cite{Martinez2019}.

\subsubsection{CTFs}
The SSB CTF is related to the geometric size of the double disk overlaps for each spatial frequency $Q_p$ for a given convergence angle $\alpha$ ~\cite{YANG20152} as
\begin{align*}
CTF(Q_p) = & \frac{4}{\pi}\cdot \left[ cos^{-1}\left(\frac{Q_p}{2}\right)-\frac{Q_p}{2}\cdot \sqrt{1-\left(\frac{Q_p}{2}\right)^2}-\right. \\
         & \left. cos^{-1}(Q_p)+Q_p\cdot\sqrt{1-\left(Q_p\right)^2}\right]
\end{align*}
for $0 \leq Q_p \leq \alpha$ and
$$ CTF(Q_p) = \frac{4}{\pi}\cdot \left[cos^{-1}\left(\frac{Q_p}{2}\right)-\frac{Q_p}{2}\cdot\sqrt{1-\left(\frac{Q_p}{2}\right)^2} \right]$$
for $\alpha <Q_p \leq 2~\alpha$ and 0 otherwise. The 2D CTF of SSB is shown in Fig.~\ref{progress}c and the real spaced transformation of this, the PSF, in panel d, which in this paper, we refer to as the SSB CTF based kernel.

Note that the PSF (Fig.~\ref{progress}d), the inverse Fourier transform of the CTF, is a result of coherent interference and contains negative values. In contrast, the ADF-based PSF is obtained via Zernike polynomials and the result of the inverse Fourier transform is squared as a result of the incoherent image formation process. We further note that an alternative to an analytical calculation of the CTF  is to extract the CTF numerically from simulations of an isolated atom via the same algorithm as used for the target image. This is indeed possible as the exact profile is negligibly dependent on the element~\cite{OLEARY2021}. 

  \begin{figure*}[htbp]
 \centering\includegraphics[width=0.95\textwidth]{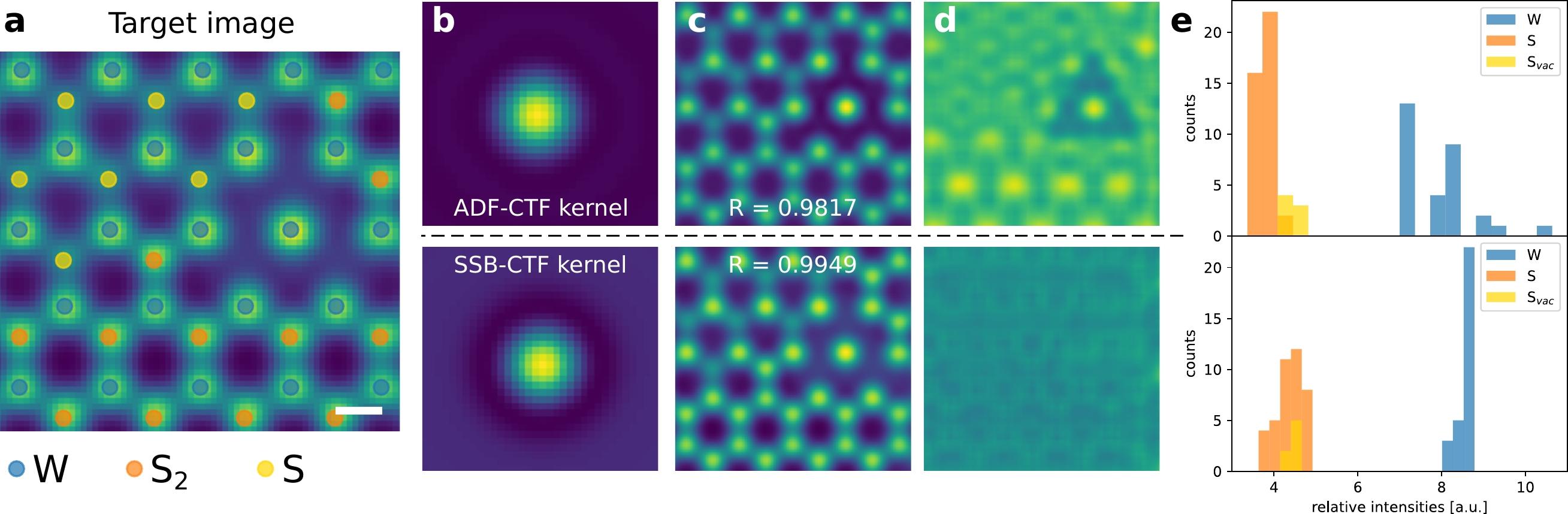}%
 \caption{\label{opt} \textbf{Comparison between phase extraction using a Gaussian kernel (based on an ADF-CTF) and a SSB-CTF based kernel}.a) Target image which is used as benchmark for the methods. b) ADF-CTF based and SSB-CTF based kernel (top and bottom, respectively). c) Optimized convolution simulation using the ADF-based kernel and SSB-based kernel in top and bottom, respectively. d) Difference image between (a) and (c). e) Extracted phases using both kernels.}
 \end{figure*}
\subsubsection{Optimization}
To quantify the merit of a particular configuration we use the correlation between the input target image and the convolutional simulation 
\begin{align}
R = \sum\limits_{i=1}^N\frac{\left( \mu^{\mathrm{sim}}-\phi^{\mathrm{sim}}_i\right) \left(\mu^{\mathrm{target}}-\phi^{\mathrm{target}}_i\right) }{\sigma^{\mathrm{sim}}\sigma^{\mathrm{target}}\left(N-1\right)}, 
\end{align}
with $\mu$ being the mean value and $\sigma$ being the  standard deviation. The sum iterates through all pixel values $\phi$ of the convolutional simulation and the input target image. This is shown in Fig.~\ref{progress}e for the initial and optimized simulation in top and bottom, respectively. The correlation function is maximized by adjusting the parameters for the atomic positions, width of the kernel and the strength of the potential. The difference images for both cases are shown in (f). The correlation in the course of the optimization is shown in (g). This function is maximized based on quadratic interpolation where the gradients in each direction are calculated by the finite difference method and the minimum of a quadratic fit is estimated to be the next iteration value. The source size can also be optimized, and the atomic positions collectively changed by applying translations, scalings, rotations and importantly sample tilts. This is crucial for experimental conditions, where a sufficient alignment between the simulation of the model and the experimental image is the task of the optimization.

  \begin{figure*}[htbp]
 \center\includegraphics[width=0.95\textwidth]{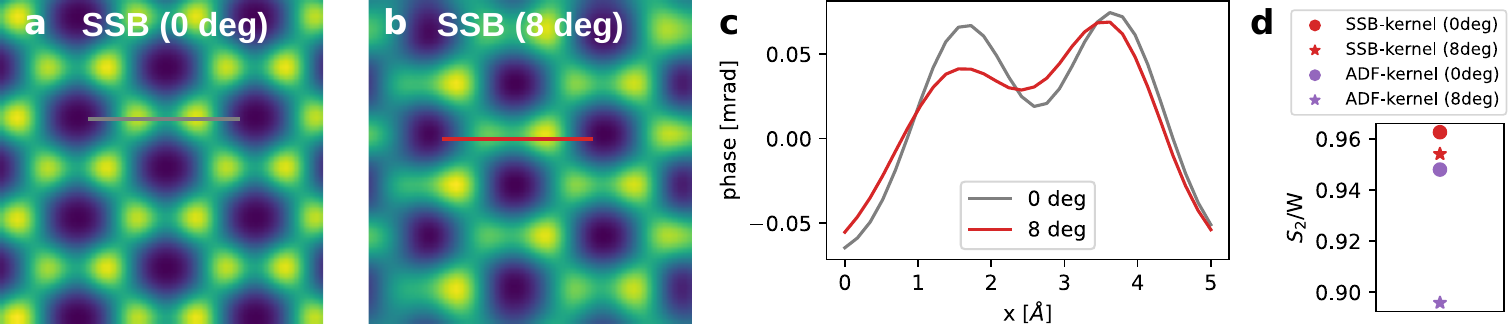}%
 \caption{\label{tilted} \textbf{Specimen tilt sensitivity of ptychographic reconstructions.} a) SSB image of WS\textsubscript{2}. b) 8 degree tilted SSB image. c) Line profile of a) and b). d) Extracted phases using a Gaussian kernel (purple) and a SSB-CTF based kernel (red). }
 \end{figure*}
\section{Results and Discussion}
To compare a conventional integration of the atomic sites with the new method, we performed the optimization analysis with an ADF-CTF based kernel and an SSB-CTF based kernel. The target input image is a quantitative SSB simulation of defective WS\textsubscript{2} (Fig.~\ref{opt}a), where the positions are labeled with blue (W), yellow (S vacancies) and orange (S\textsubscript{2}). The input simulation is based on an independent atomic model potential, in which electron redistribution is not taken into account. The convergence angle is set to 20~mrad and the acceleration voltage is 60~kV. Therefore, the same elements have the same phase in the phase image which is an ideal test for the different methods. The ADF-CTF based kernel is very similar to a Gaussian kernel and the analysis with it is very similar to simple Gaussian-fitting~\cite{VANAERT2009} or Voronoi-integration~\cite{E2013} in the absence of aberrations~\cite{HOFER2021}. However, we will show that this approach is problematic for quantifying SSB images as the phase of the atoms also depends on the local environment, as discussed in the introduction. The second kernel is obtained via the SSB-CTF, as described in the Methods section. Crucial for an accurate analysis is the negative halo in the kernel, which is also visible in the SSB reconstruction as shown in Fig.~\ref{distance}a. Both kernels are shown in Fig.~\ref{opt}b.

\begin{figure}[h]
 \centering\includegraphics[width=0.45\textwidth]{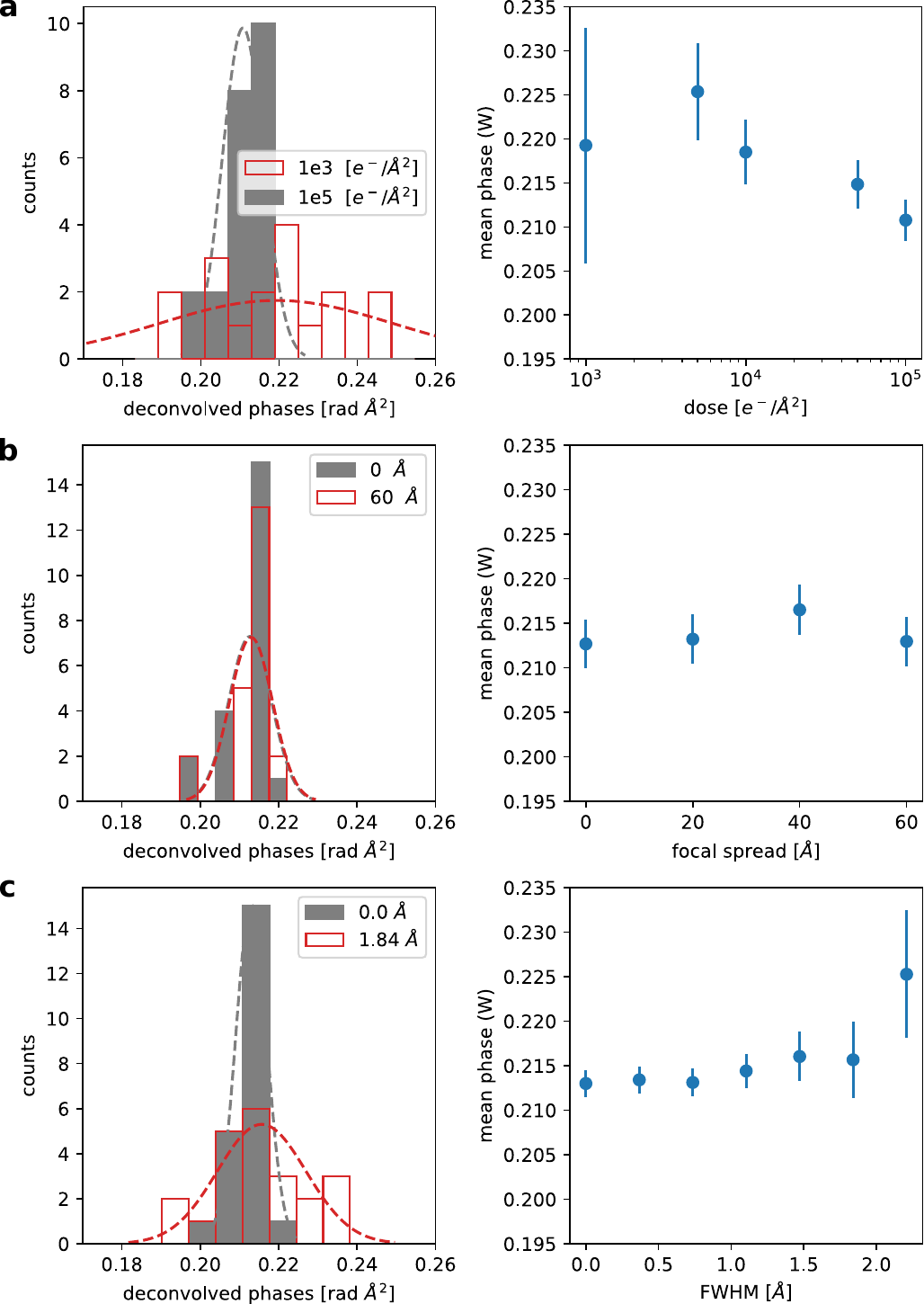}%
 \caption{\label{sigma} \textbf{Precision of optimization method at different conditions.}  Analysis of SSB phase extraction with different electron doses (a), with and without temporal incoherence (b) and with and without the effect of finite source sizes (c). Each panel shows the extracted phases of two conditions left and the mean values as well as the error bars right.  }
 \end{figure}
When optimizing the simulations using the atomic positions and point-potential values as parameters, both methods result in similar simulated images (cf.\ Fig.~\ref{opt}c). The difference images between the optimized convolutional simulations and the input target image, however, shows some residual features in the case of the ADF-CTF based image (cf.\ panel d). This is because the contrast cannot be perfectly matched in this case. The difference image using the SSB-CTF kernel is almost perfectly flat as a result of an excellent match between the simulation and the target image. This is reflected in the higher correlation coefficient in the SSB-CTF kernel based optimization. 

More importantly, significant errors appear in the resulting extracted phases of the optimized point-potentials. When using the ADF-based kernel simulation, the histogram of phases (Fig.~\ref{opt}e) shows a more scattered W phase value distribution as the stronger phase of  W close to weaker potentials (e.g.\ vacancies) has to be matched by increasing the point-potential. The highest peak corresponds to the isolated W, surrounded by double vacancies in the structure, due to the contrast reduction for the W atoms within the pristine lattice. Also, the S atoms in the S\textsubscript{2} sites (orange) are slightly lower than the single S vacancies (yellow) due to the same reason (see Fig~\ref{opt}e top). This makes the phase extraction unreliable as the contrast transfer functions are not properly taken into account. Using the SSB-CTF kernel gives, however, a much sharper histogram as the local environment is taken into account. Again, keep in mind that this is an independent atom model so the variation should be essentially nil. Correctly accounting for the CTF makes the interpretation of the phases due to actual changes in the material, such as bonding, much easier and more reliable. Note that the width of S (orange) in the S\textsubscript{2} site is unavoidably larger because they are pairwise overlapping in the projection. However, their mean value is matching the S vacancy values (yellow) as expected for the target image. Also the W atoms show much less variation despite the non-linear contrast formation. This is important for detecting small differences between atomic sites, which occurs for example when detecting charge transfer.

The strong dependency of the SSB contrast on the local environment also makes quantification difficult when small sample tilts are present as these lead to small projected positional shifts of the atoms. Small tilts are difficult to measure in STEM experiments and residual sample tilts are practically unavoidable. The contrast dependency of the sample tilt is demonstrated in Fig.~\ref{tilted}a and b, which show IAM-based SSB simulations of an untilted and 8 degrees tilted WS\textsubscript{2} respectively. The contrast change is shown by the line profile (panel c), where the S\textsubscript{2} sites show a significantly reduced phase when tilted. To demonstrate the robustness of the optimization method to sample tilt, we performed the quantification analysis on both the untilted and tilted images using both the SSB and ADF based kernels. %We again extracted the phases of the optimized point potential. 
For the untilted pristine case, both kernels give similar S\textsubscript{2} to W ratios. However, analysing the tilted image gives a remarkable deviation when using the ADF-kernel (5.50\%) while the difference using the SSB-CTF based kernel is much lower (0.89\%). Ideally mistilt should be avoided experimentally, but this method can greatly reduce the influence of mistilt, making it potentially insignificant compared to other influences such as limited electron dose. This shows the robustness of the new phase extraction method to sample tilt. To make this happen, the sample tilt is included in the optimization which gives the correct value based on the distortion of the lattice. Importantly, this requires that scan distortion is minimized, which is now possible in 4D STEM based on truly fast cameras that enable microsecond level dwell times, such as event-based cameras~\cite{JANNIS2022}.

Next we investigated the robustness of the phase extraction to other typical experimental influences. We measured the W phase as a function of electron dose, partial coherence and source size individually. The resulting phases are shown in Fig.~\ref{sigma}. For common operating conditions, by far the largest deleterious influence on the precision is when a limited amount of signal per diffraction pattern is available, as a result of a limited electron dose imparted to the specimen. This is shown in Fig.~\ref{sigma}a. However the error bar of the mean value not only depends on the dose but also on the number of atomic sites analysed. To evaluate if the error bar is small enough in the context of analysing phase differences, for example due to the subtle changes occurring with charge transfer due to bonding, the difference between the involved phases has to be considered. We analysed the charge transfer of WS\textsubscript{2} where the phase difference was indeed large enough to detect the difference caused by charge transfer at moderate doses~\cite{hofer2023detecting}. The partial coherence is simulated via a defocus spread ranging from 0--6~nm. Typical values are around 3~nm for a cold FEG, greater for less coherent sources. The phase change is small because of the robustness of ptychography to partial coherence~\cite{PENNYCOOK2019}. Lastly, the source size is simulated by using a Gaussian blur up to a FW 2.3~\AA. The phase change is insignificant for typical source sizes~\cite{VERBEECK201235}. We therefore conclude that dose and therefore the sample stability is the most crucial parameter to consider if phase changes are detectable. 

Phase retrieval methods can encounter the issue of phase wraps with strong potentials. This means when increasing the strength of the potential, the resulting atomic phase wraps to a negative value in the center of the atomic columns as has been observed in both simulations and experiment. At this point, the atoms deviate from their natural rotational symmetric shape and the presented method is not reliable.  However, it has been shown that these effects can be corrected by adjusting the probe defocus post  collection~\cite{Chuang}. However, as the ptychographic contrast mechanism becomes more complex when the sample gets thicker we consider the present method to be reliable for thin materials such as 2D materials. The application of this method to thicker materials is beyond the scope of this paper and will be investigated in future work.

 \begin{figure}[ht]
 \center\includegraphics[width=0.45\textwidth]{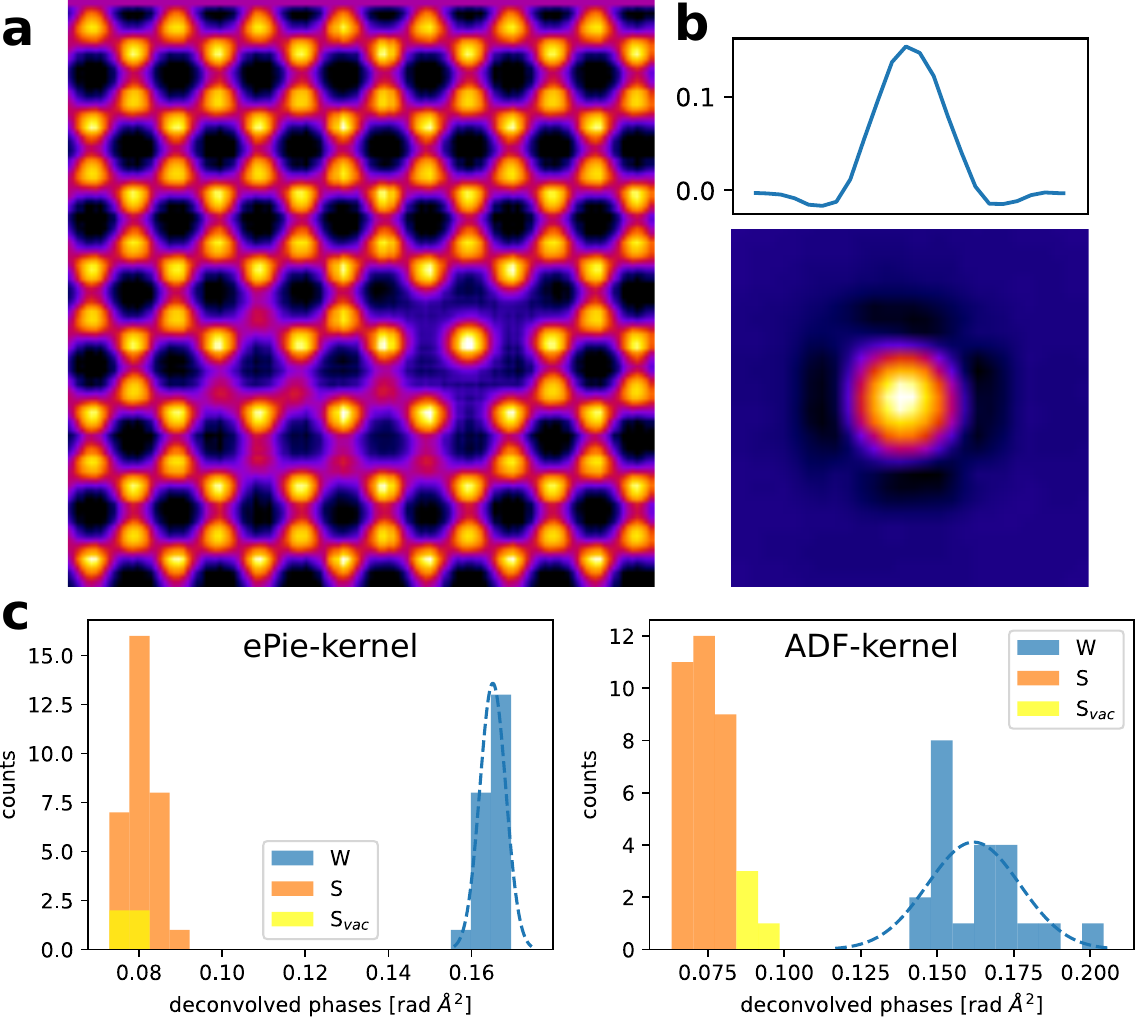}%
 \caption{\label{epie} \textbf{Quantification of iterative ptychography using ePIE.} a) ePIE reconstruction of defective WS\textsubscript{2}. b) Line profile and 2D image of PSF of ePIE. c) Extracted phases of (a) using the ePIE kernel (left) and a Gaussian kernel (right). ePIE also exhibits a PSF with a negative halo which enhances the phase of an isolated atom. This can be correctly accounted for using the ePIE PSF with our method in quantification.}
 \end{figure}

 Our method is in principle also applicable to imaging methods with other CTFs. The choice of CTF results in different kernel shapes representing the shape of an individual atom in the final phase image~\cite{OLEARY2021}. This makes the method very flexible to use, including for noise-normalized contrast transfer functions. Here, we analyse the phase retrieved by iterative ptychography with the extended ptychographic iterative engine (ePIE) as an example. We simulated the same region as the other examples via ePIE which has a different CTF and therefore a different PSF than SSB, as shown in Fig.~\ref{epie}a. Since the exact CTF depends on the algorithm, we provided the kernel by simulating a single atom with the same algorithm. We used 0.5~\AA  as the probe position step size, a 1~nm defocus and the same 20~mrad convergence angle used earlier. Interestingly, it also shows a negative halo, similar to the SSB, but somewhat less pronounced (Fig.~\ref{epie}b). Therefore, the W within the defective network is also brighter than the W atoms within the pristine lattice with ePIE. As with SSB data, the kernel method again correctly accounts for this when using the ePIE-based kernel. The extracted phases show, similar to the SSB, sharp peaks in the case of the optimization using the ePIE-PSF. Using the Gaussian-shaped PSF from the ADF kernel on the ePIE data, results in a much broader width of the distribution of W phases, as expected (cf.\ Fig.~\ref{epie}c).

\begin{figure}[ht]
 \center\includegraphics[width=0.45\textwidth]{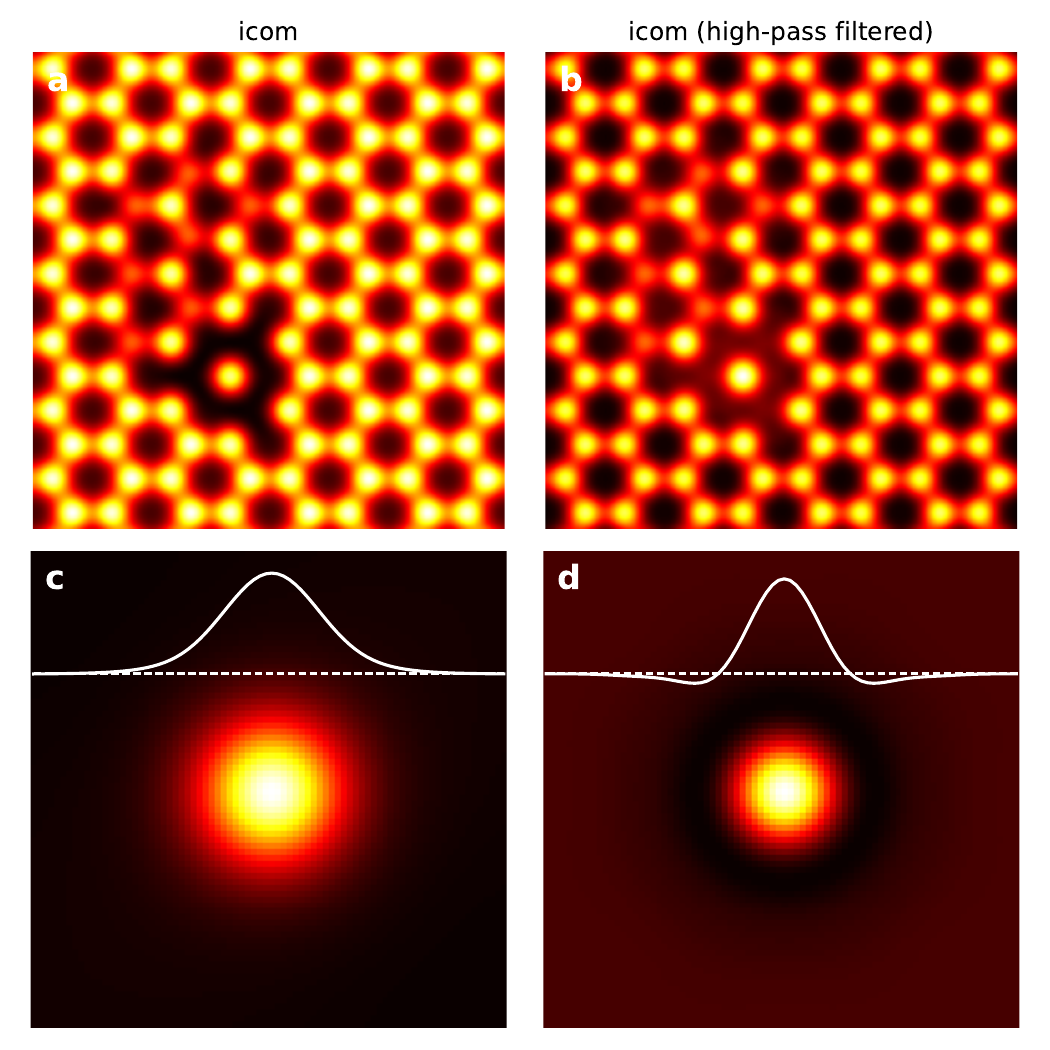}%
 \caption{\label{icom} \textbf{Filter effect on phase quantification of iCoM} a) Raw iCoM reconstruction of defective WS\textsubscript{2}. b) Same as (a) but with a high-pass filter. c,d) PSF of (a) and (b), respectively. Notice that the filter causes a negative halo in the PSF of the iCoM which again enhances the brightness of isolated atoms compared to those with closer neighbors. This can be accounted for with our method.}
 \end{figure}

 We also analysed integrated center-of-mass (iCoM) imaging as a further 4D STEM imaging method. iCoM is becoming increasingly popular with the development of fast cameras since it is a more accurate method compared to the now popular integrated differential phase contrast (iDPC) imaging modality. However, both images can be analysed similarly as outlined below. The only difference is the non-rotational symmetric CTF of iDPC as a result of the segmentation of the detector, which would need to be accounted for correctly for an accurate analysis of iDPC images, and therefore the exact detector geometry with respect to the diffraction patterns would have to be known. iCoM is very simple and fast and therefore a reliable tool for live iCoM imaging~\cite{ricom}. However, it has been observed to be less dose efficient than ptychographic methods~\cite{Chuang,Leary2020}. It also cannot correct residual aberrations and has no super-resolution ability. Since the CTF of iCoM is starting at 1~\cite{LAZIC2016265}, there are no artifacts associated with the negative part of the PSF when used with the raw data. However, high pass filters are commonly applied to iCoM and iDPC data, and riCoM contains an intrinsic filtering based depending on the riCoM kernel used.  In all these cases the PSF for iCoM, and by extension iDPC, obtains a negative contribution similar to the discussions above. This is demonstrated in Fig.~\ref{icom} where raw and high pass filtered iCoM images of the same defective WS\textsubscript{2} region are shown. Such filtering is very commonly used to sharpen iCoM or iDPC images generally or ameliorate the effect of contamination, as well as edge artifacts. Importantly, the isolated W atom is also brighter here in the case of the filtered image. This is, once again, a result of the PSF which shows a negative tail in case of the filtered image. Therefore, conventional quantification of high-pass filtered (iCoM) images will encounter similar problems as in the SSB image analysis in that the extracted phases highly depend on the environmental conditions. Omitting the high pass filter however also often leads to unreliable quantification as the low-frequency contribution harms the image analysis~\cite{Yücelen2018}. 
Therefore, a reliable quantification of iCoM and iDPC phases is only possible via a method such as the optimization method we have presented in which the exact filter conditions are reproduced in the PSF. This is usually facile since the filter parameters are usually very accurately known.

 \section{Conclusion}
 We have introduced a new method to reliably analyse ptychographic phase images quantitatively. The method takes into account the specific contrast transfer function and is therefore robust to local environments such as atomic configuration and sample tilts. This is very important when tiny changes of the phase images are analysed, such as with charge transfer due to bonding. We believe this method opens a path towards a reliable quantification of phase images as it has been achieved for ADF images.

 \section{Acknowledgement}
We acknowledge funding from the European Research Council (ERC) under the European Union's Horizon 2020 Research and Innovation Programme via Grant Agreement No. 802123-HDEM and FWO Project G013122N “Advancing 4D STEM for atomic scale structure property correlation in 2D materials” (C.H.). 

\bibliography{references}

\end{document}